\documentclass[aps,floats]{revtex4}
\usepackage{amsmath,amssymb}
\usepackage{graphicx,epsfig}

\begin{document}
\bibliographystyle {plain}

\def\oppropto{\mathop{\propto}} 
\def\opsimeq{\mathop{\simeq}}
\def\opoverderline{\mathop{\overline}}
\def\operarrow{\mathop{\longrightarrow}}
\def\opsim{\mathop{\sim}}

\def\fig#1#2{\includegraphics[height=#1]{#2}}
\def\figx#1#2{\includegraphics[width=#1]{#2}}


\title{ Random Transverse Field Ising Model in dimension $d=2,3$ : \\
Infinite Disorder scaling via a non-linear transfer approach   } 


\author{ C\'ecile Monthus and Thomas Garel }
 \affiliation{Institut de Physique Th\'{e}orique, CNRS and CEA Saclay
91191 Gif-sur-Yvette cedex, France}

\begin{abstract}

The 'Cavity-Mean-Field' approximation developed for the Random Transverse Field Ising Model
on the Cayley tree [L. Ioffe and M. M\'ezard, PRL 105, 037001 (2010)] has been found
 to reproduce the known exact result for the surface magnetization in $d=1$
[O. Dimitrova and M. M\'ezard, J. Stat. Mech. (2011) P01020]. In the present paper,
we propose to extend these ideas in finite dimensions $d>1$
via a non-linear transfer approach for the surface magnetization.
In the disordered phase, the linearization of the transfer equations
correspond to the transfer matrix for a Directed Polymer in a random medium
of transverse dimension $D=d-1$, in agreement with the leading order perturbative
scaling analysis [C. Monthus and T. Garel, arxiv:1110.3145]. We present
numerical results of the non-linear transfer approach in dimensions $d=2$
and $d=3$. In both cases, we find that the critical point is
governed by Infinite Disorder scaling. In particular exactly at criticality,
the one-point surface magnetization scales as $\ln m_L^{surf} \simeq - L^{\omega_c} v$,
where $\omega_c(d)$ coincides with the droplet exponent $\omega_{DP}(D=d-1)$
of the corresponding Directed Polymer model, with
 $\omega_c(d=2)=1/3$ and $\omega_c(d=3) \simeq 0.24$. The distribution $P(v)$ 
of the positive random variable $v$ of order $O(1)$
presents a power-law singularity near the origin $P(v) \propto v^a$ with $a(d=2,3)>0$ so that all
moments of the surface magnetization are governed by the same power-law decay
$\overline{ (m_L^{surf})^k } \propto L^{- x_s}$ with $x_s=\omega_c (1+a)$ 
independently of the order $k$.

\end{abstract}

\maketitle

\section{ Introduction }

The quantum Ising model 
\begin{eqnarray}
{\cal H} =  -  \sum_{<i,j>} J_{i,j}  \sigma^z_i \sigma^z_j - \sum_i h_i \sigma^x_i
\label{hdes}
\end{eqnarray}
where the nearest-neighbor couplings $J_{i,j}>0$
and the transverse-fields $h_i>0$ are independent random variables
drawn with two distributions $\pi_{coupling}(J)$ and $\pi_{field}(h)$
is the basic model to study quantum phase transitions at zero-temperature
in the presence of frozen disorder.
In dimension $d=1$, exact results for a large number of observables
have been obtained by Daniel Fisher \cite{fisher} 
 via the asymptotically exact Strong Disorder renormalization procedure
 (for a review, see \cite{review_strong}). In particular, the transition is governed
by an Infinite Disorder fixed point 
and presents unconventional scaling laws with respect to the pure case.
In dimension $d>1$, the Strong Disorder renormalization procedure can still be defined.
It cannot be solved analytically, because the topology of the lattice changes upon renormalization,
but it has been studied numerically with the conclusion that the transition is also governed by
an Infinite Disorder fixed point in dimensions $d=2,3,4$ 
 \cite{motrunich,fisherreview,lin,karevski,lin07,yu,kovacsstrip,
kovacs2d,kovacs3d,kovacsentropy,kovacsreview}.
These numerical renormalization results
 are in agreement with the results of independent quantum Monte-Carlo
in $d=2$ \cite{pich,rieger}.

Even if it is clear that the most natural method to study Infinite Disorder fixed points
is the Strong Disorder renormalization approach, it seems useful to determine whether
other approaches are able to describe Infinite Disorder scaling. 
In this paper, we introduce a simple non-linear transfer approximation
 for the surface magnetization 
in finite dimension $d>1$, which is inspired from the
 'Cavity-Mean-Field' approximation
developed in Refs  \cite{ioffe,feigelman,dimitrova}, 
and we study numerically the critical properties
of this approximation in dimensions $d=2$ and $d=3$,

The paper is organized as follows.
In Section \ref{sec_transfert}, we recall briefly the
 'Cavity-Mean-Field' approximation
developed in Refs  \cite{ioffe,feigelman,dimitrova}
and introduce the non-linear transfer approach for finite dimensions $d>1$.
Our numerical results in dimension $d=2$ and $d=3$ are presented in sections \ref{sec_dim2}
and \ref{sec_dim3} respectively.
Our conclusions are summarized in section \ref{sec_conclusion}.

\section{ Non-linear transfer approach for the surface magnetization  }

\label{sec_transfert}

\subsection{ 'Cavity-Mean-Field' approximation on the Cayley tree \cite{ioffe,feigelman,dimitrova} }

For the random quantum Ising model model defined on a tree of coordinence $(K+1)$,
the following 'Cavity-Mean-Field' approximation has been developed \cite{ioffe,feigelman,dimitrova} :
an ancestor $i$ is submitted to the effective single spin Hamiltonian
\begin{eqnarray}
H_i^{eff} =  - B_i \sigma^z_i  - h_i  \sigma^x_i  
\label{heffbihi}
\end{eqnarray}
where $h_i$ is its own random transverse field, and where $B_i$
represents the longitudinal field created by the $K$ children $j$ (related to $i$ by the
ferromagnetic couplings $J_{ij}$)
within a 'Mean-Field approximation' ( the operator $\sigma^z_j $ is replaced by its expectation value
$< \sigma^z_j > $)
\begin{eqnarray}
B_i=\sum_{j=1}^K J_{i,j} < \sigma^z_j >
\label{bimf}
\end{eqnarray}
The effective Hamiltonian of Eq. \ref{heffbihi} is only a two-level system
that can be solved immediately : the magnetization of the ground state reads
\begin{eqnarray}
m_i \equiv <\sigma_i^z>_{H_i^{eff} } =  \frac{B_i}{\sqrt{B_i^2+h_i^2}}
\label{miheff}
\end{eqnarray}
Using Eq. \ref{bimf}, one obtains the following non-linear recurrence
 for the magnetizations $m_i$
(see Eq. 4 of \cite{ioffe}, Eq. 7 of \cite{feigelman}, Eq. 17 
of \cite{dimitrova} in the limit of zero temperature $\beta=+\infty$)
\begin{eqnarray}
m_i  =  \frac{ \sum_{j=1}^K J_{i,j} m_j }{\sqrt{ \left( \sum_{j=1}^K J_{i,j} m_j \right)^2+h_i^2}}
\label{mimj}
\end{eqnarray}
We refer to Refs \cite{ioffe,feigelman,dimitrova} for more details on
 this 'Cavity-Mean-Field'
approximation and on its properties.
As a final remark, let us stress that the 'Cavity-Mean-Field' is not exact for the pure model
on the Cayley tree (see Fig. 3 of Ref \cite{dimitrova}), but has been argued to become
quantitatively correct in the limit of high connectivity $K \gg 1$  \cite{ioffe,feigelman,dimitrova}.

In the disordered phase where the magnetizations flows towards zero, 
the  non-linear recurrence of Eq. \ref{mimj} can be linearized to give the following
recursion 
\begin{eqnarray}
m_i  \simeq  \frac{1}{h_i} \sum_{j=1}^K J_{i,j} m_j 
\label{mimjlinear}
\end{eqnarray}
which is equivalent to the problem of a Directed Polymer on the Cayley tree
 \cite{ioffe,feigelman,dimitrova}. This equivalence can be justified directly
at the level of lowest-order perturbation theory 
(i.e. without invoking the 'Cavity-Mean-Field' approximation of Eq. \ref{mimj}),
and can be in this way extended to the finite-dimensional case \cite{transverseDP}.

\subsection{ 'Cavity-Mean-Field' approximation in $d=1$ \cite{dimitrova} }

\label{Cavity1d}

For $K=1$, the Cayley tree of coordinence $(K+1)$ discussed 
in the previous section
becomes a one-dimensional chain, and Eq. \ref{mimj} becomes the
 one-dimensional non-linear
recurrence \cite{dimitrova}
\begin{eqnarray}
m_i  =  \frac{ J_{i,i+1} m_{i+1} }{\sqrt{ \left( J_{i,i+1} m_{i+1} \right)^2+h_i^2}}
\label{mimjd1}
\end{eqnarray}
Assuming one starts with the boundary condition $m_L=1$ at site $i=L$, 
one obtains
the following explicit expression for the surface magnetization
 $m_0^{surf}$ at the site $i=0$
\cite{dimitrova}
\begin{eqnarray}
m_0^{surf}= 
\left[ 1+ \sum_{i=0}^{L-1} \prod_{j=0}^i \left( \frac{h_j}{J_{j,j+1} } \right)^2 \right]^{-1/2}
\label{msurfexact}
\end{eqnarray}
As stressed in \cite{dimitrova}, this expression exactly coincides
 with the rigorous expression 
that can be obtained from a free-fermion representation \cite{peschel,msurf},
and from which many critical exponents can be obtained 
\cite{msurf,dhar,ckesten}.

The reason why the 'Cavity-Mean-Field' approximation turns out to become
 exact for the surface magnetization in $d=1$ is not clear to us, and seems rather surprising :
usually 'mean-field approximation' are exact in sufficiently high dimensions or on trees, 
and are not exact in low dimensions,
the 'worst case' being precisely $d=1$. 
Here we have exactly the opposite conclusion : 
the 'Cavity-Mean-Field' is not exact for the pure model on the tree (for the disordered case,
it is not known), but turns out to be exact in $d=1$, both for the pure and the disordered case.
In the absence of any satisfactory explanation for this unusual situation,
we tend to think that the exactness of the 'Cavity-Mean-Field' in $d=1$
is likely to be a 'coincidence' specific to this particular case,
from which one cannot draw general conclusions for the validity of this approach in higher dimensions
or for other quantum disordered models.

\subsection{ Non-linear transfer approach in finite dimension $d>1$ }

\subsubsection{ Description }

\begin{figure}[htbp]
 \includegraphics[height=8cm]{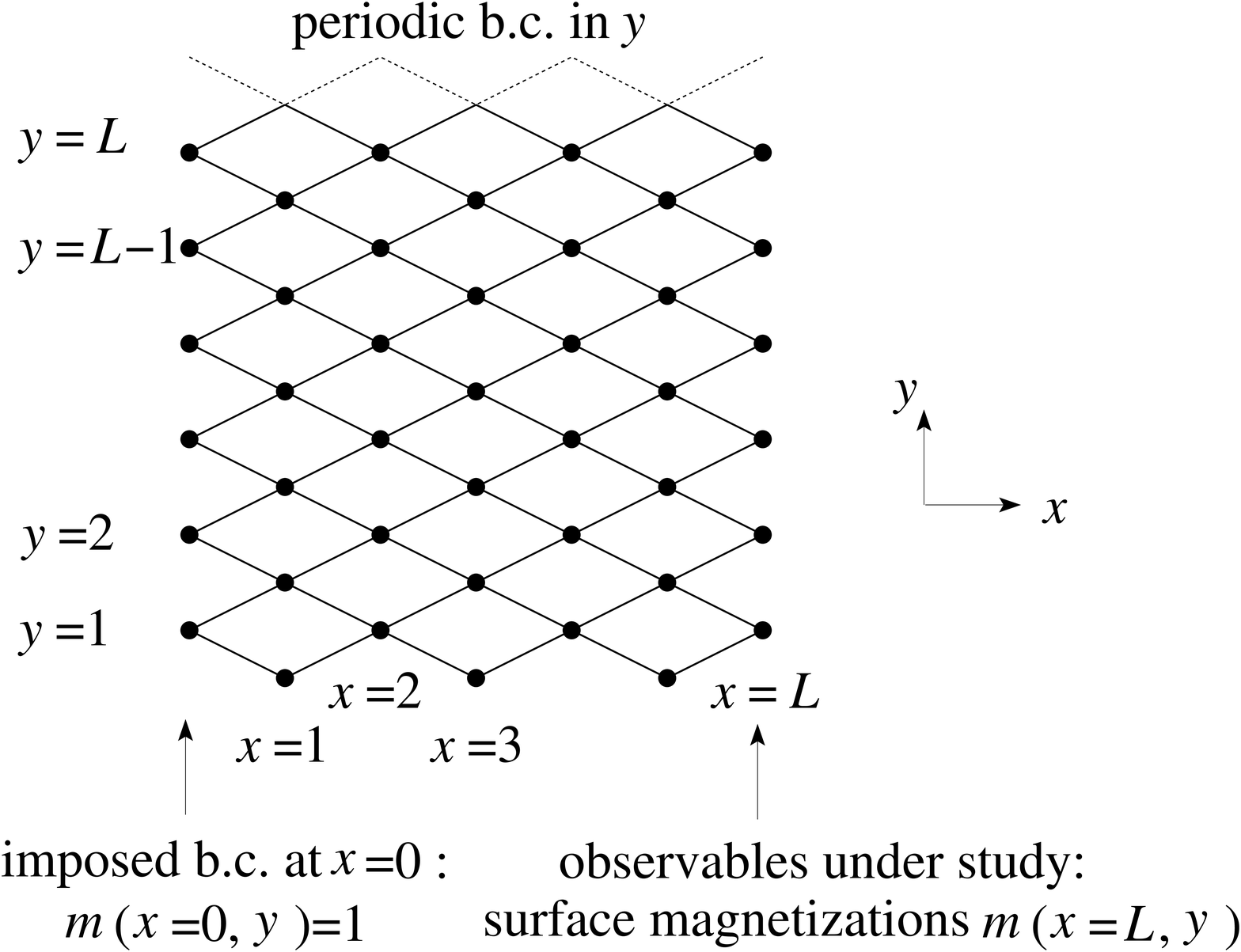}
\caption{ Notations to define the non-linear transfer approach in $d=2$ :
we impose the boundary conditions $m(x=0,y)=1$ on the left boundary,
and we study the surface magnetizations $m(x=L,y)=1$
 on the right boundary (see text for more details). }
\label{figgeom2d}
\end{figure}

In finite dimensions $d>1$, the authors of Ref \cite{dimitrova} 
have proposed to use the 'Cavity-Mean-Field' approximation on a Cayley tree
with parameter $K=2d-1$ (to reproduce the connectivity of each spin). 
In the present paper, we propose instead to extend Eqs \ref{mimj}
towards an appropriate non-linear transfer approach
 for the surface magnetizations of a finite sample of volume $L^d$.

For clarity, let us first explain the procedure for the case $d=2$.
As shown on Fig. \ref{figgeom2d}, we consider a lattice containing $L^2$ 
spins : when $x$ is even ($x=0,2,4,..$), the coordinate $y$ takes the $L$ 
integer-values $y=1,2,...,L$ ;
when $x$ is odd ($x=1,3,...$), the coordinate $y$ takes the $L$ 
half-integer-values $y=3/2,5/2,...,L+1/2$.
The boundary conditions are periodic in $y$ with $y+L \equiv y$.
At $x=0$, we impose the boundary condition of unity magnetization
\begin{eqnarray}
m (x=0,y) =  1
\label{mix0}
\end{eqnarray}
and we are interested in the $L$ surface magnetizations $m(x=L,y)$
at the opposite boundary $x=L$.

For this situation, we propose to use the ideas of Eqs \ref{mimj}
within the following transfer approach. 
We assume that we have already found the surface magnetizations 
 on the column $m(x-1,y)$, and we add another column of $L$ sites at $x$.
From the Cavity point of view, the new spin at $(x,y)$ is submitted
to its own  random transverse field $h(x,y)$ and 
to the longitudinal field (see Eq. \ref{bimf}) 
created by its two neighbors on the column $x-1$
\begin{eqnarray}
B(x,y)= 
J_{\{\left(x,y\right), \left(x-1,y+\frac{1}{2}\right)\}}
 m \left(x-1,y+\frac{1}{2}\right)
+J_{\{\left(x,y\right), \left(x-1,y-\frac{1}{2}\right)\}}
 m \left(x-1,y-\frac{1}{2}\right)
\label{bimf2d}
\end{eqnarray}
so that its surface magnetization reads (Eq. \ref{miheff})
\begin{eqnarray}
m (x,y) =   \frac{B(x,y)}{\sqrt{B^2(x,y)+h^2(x,y)}}
\label{miheff2d}
\end{eqnarray}
Eqs \ref{bimf2d} and \ref{miheff2d} define a non-linear transfer procedure
that can be iterated from the boundary condition on the column $x=0$
 of Eq. \ref{mix0} up to $x=L$, where we analyze the statistics 
of the final surface magnetizations $m(x=L,y)$.
It is clear that the generalization of this procedure
to $d=3$ is straightforward : we add another direction $z$
with periodic boundary conditions that plays exactly the same role as $y$.

\subsubsection{ Linearized transfer matrix within the disordered phase }

Within the disordered phase, the surface magnetizations $m(x=L,y)$
are expected to decay typically exponentially in $L$, so that 
one may linearize the transfer Eqs \ref{bimf2d} and \ref{miheff2d}
to obtain
\begin{eqnarray}
{ \bf Linearization : \ \ } m (x,y) \simeq  
\frac{J_{\{\left(x,y\right), \left(x-1,y+\frac{1}{2}\right)\}}}{h(x,y)}
 m \left(x-1,y+\frac{1}{2}\right)
+ \frac{J_{\{\left(x,y\right), \left(x-1,y-\frac{1}{2}\right)\}}}{h(x,y)}
 m \left(x-1,y-\frac{1}{2}\right)
\label{milinear}
\end{eqnarray}
This linearized equations can be derived directly within a lowest-order
 perturbative approach \cite{transverseDP}
 (i.e. without invoking the 'Cavity-Mean-Field'
 approximation) and  corresponds
 to the transfer matrix satisfied by the 
partition function of a Directed Polymer 
with $D=(d-1)$ transverse directions, 
as discussed in detail in \cite{transverseDP}.
We refer to \cite{transverseDP} for the description of the consequences
of this correspondence, and for the analogy with Anderson localization, where
the droplet exponent of the Directed Polymer also appears in the localized phase
\cite{NSS,medina,prior}.
 Here our conclusion is that the
non-linear transfer approach describes at least correctly
the disordered phase, where it coincides with the lowest-order
 perturbative approach \cite{transverseDP}.

\subsubsection{ Discussion }

Besides its correctness in the disordered phase that we have
 just discussed, the validity of the non-linear transfer
exactly at criticality and in the ordered phase has to be studied
for the disordered case in $d>1$.
Since it has been found to be exact in $d=1$ (see section \ref{Cavity1d}),
one could hope that it is not 'too bad' in $d=2,3$ 
(even if it is clear that this approximation is not valid for the pure model) :
 we believe that it should capture correctly the nature of the transition
between 'Infinite-Disorder' or 'Conventional' scaling. 
In the following, we present our numerical results in $d=2$ and $d=3$
and discuss the scaling properties in the two phases and at criticality.

\section{ Numerical results in dimension $d=2$  }

\label{sec_dim2}

In this section, we present the numerical results obtained 
with the following sizes $L$ and the corresponding numbers $n_s(L)$ of disordered samples
of volume $L^2$
\begin{eqnarray}
L && = 10^3, 2. 10^3 , 3.10^3 , 4.10^3 , 5.10^3 , 6.10^3, 7.10^3, 8. 10^3 \nonumber \\
n_s(L) && = 2.10^5, 13.10^4, 65.10^3, 37.10^3, 24.10^3, 17.10^3, 13.10^3, 10^4
\label{numed2}
\end{eqnarray}
For each sample $\alpha$, we collect the $L$ values of the surface
 magnetization $m^{(\alpha)}(x=L,i)$
at the different points $i=1,2,..,L$ of the surface (see Fig. \ref{figgeom2d}).
Average values and histograms are then based on these $L \times n_s(L)$ values.

We have chosen to consider the following log-normal distribution 
for the random transverse fields $h_i>0$
\begin{eqnarray}
\pi_{LN}(h) = \frac{1}{ h \sqrt{2 \pi \sigma^2 }} e^{- \frac{(\ln h - \overline{\ln h})^2}{2 \sigma^2 }}
\label{lognormal}
\end{eqnarray}
of parameter $ \overline{\ln h}=0$ and $\sigma=1$, whereas
 the ferromagnetic couplings $J_{i,j}$ are not random
but take a single value $J$ that will be the control parameter of the quantum transition.

\subsection{ Disordered phase ($J<J_c$)}

\subsubsection{  Exponential decay of the typical surface magnetization 
$m_L^{typ} \equiv e^{\overline{ \ln m_L^{surf} }}$   }

\begin{figure}[htbp]
 \includegraphics[height=6cm]{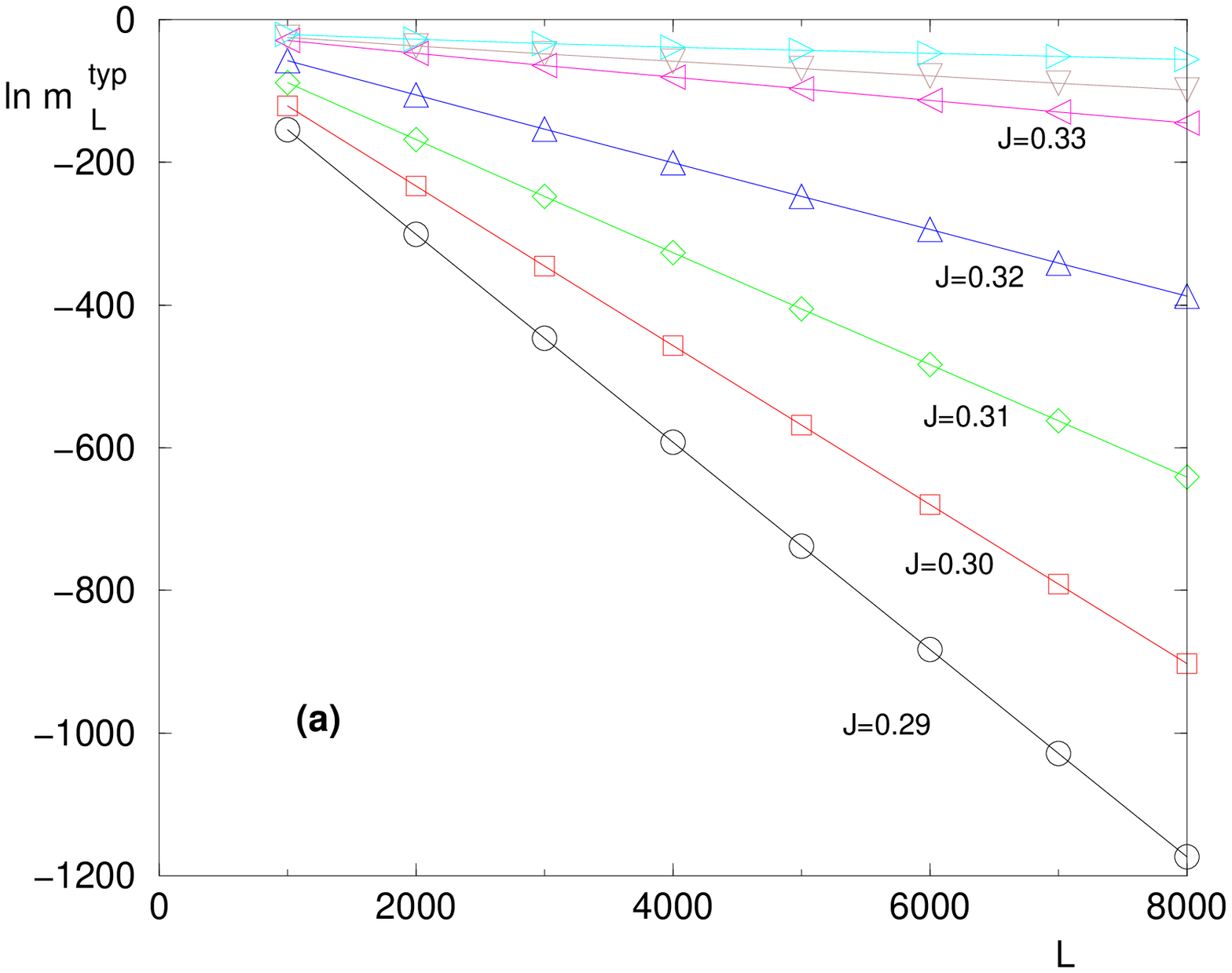}
\hspace{2cm}
\includegraphics[height=6cm]{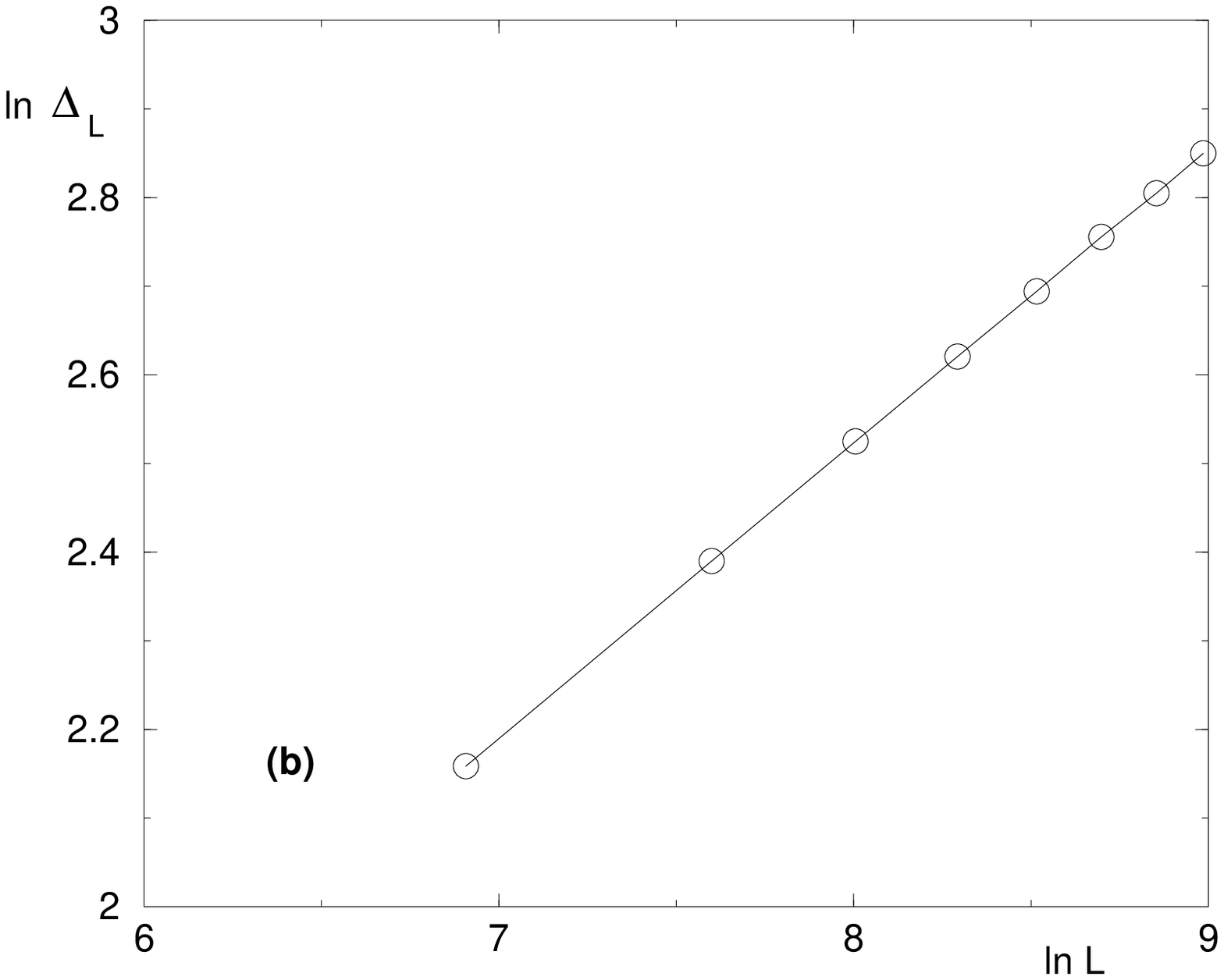}
\caption{ Disordered phase $J<J_c$ in $d=2$ 
(a) Exponential decay of the typical surface magnetization 
$m_L^{typ} \equiv e^{\overline{ \ln m_L^{surf} }}$ :
 we show the linear decay of $ \ln m_L^{typ} $ 
as a function of the length $L$  (see Eq. \ref{msdis}).
(b)  Log-log plot of the width $\Delta_L$ 
of the distribution of the logarithm of the surface magnetization
as a function of $L$ (here for $J=0.29$) : the slope is of order $\omega \simeq 0.33$
(see Eq. \ref{deltaLomega}). }
\label{fig2ddisordered}
\end{figure}

In the disordered phase $J<J_c$, one expects that the typical surface magnetization 
defined by 
\begin{eqnarray}
\ln (m_L^{typ}) \equiv  \overline{ \ln m_L^{surf} }
\label{defmstyp}
\end{eqnarray}
 decays exponentially
 with $L$
\begin{eqnarray}
\ln (m_L^{typ}) \equiv  \overline{ \ln m_L^{surf} }(J<J_c) 
\opsimeq_{L \to \infty} - \frac{L}{\xi_{typ}(J)}
\label{msdis}
\end{eqnarray}
where $\xi_{typ}$ represents the typical correlation length that diverges
at the transition as a power-law
\begin{eqnarray}
\xi_{typ}(J) \opsimeq_{J \to J_c^-} (J_c-J)^{-\nu_{typ}}
\label{xityp}
\end{eqnarray}

On Fig. \ref{fig2ddisordered} (a) we show our numerical results : 
concerning the exponential decay with $L$ of Eq. \ref{msdis} 
  for various values of $J$.
We find that the corresponding slope $1/\xi_{typ}(J)$  
vanishes  near the critical value $J_c \simeq 0.335$ with the exponent
\begin{eqnarray}
\nu_{typ} \simeq 1
\label{nutyp2d}
\end{eqnarray}

\subsubsection{ Growth of the width of the distribution of the logarithm of the surface magnetization  }

In the disordered phase $J<J_c$, one expects that the width $\Delta_L$ 
of the distribution of the logarithm of the surface magnetization defined by
\begin{eqnarray}
\Delta_L  \equiv  \left( \overline{ (\ln m_L^{surf})^2 } - (\overline{ \ln m_L^{surf} })^2 \right)^{1/2}
\label{defdeltaL}
\end{eqnarray}
grows as a power-law of $L$
\begin{eqnarray}
\Delta_L (J<J_c)  \opsimeq_{L \to \infty} L^{\omega}
\label{deltaLomega}
\end{eqnarray}

Our numerical data shown on Fig. \ref{fig2ddisordered} (b)
correspond to the value
\begin{eqnarray}
\omega(d=2) \simeq 0.33
\label{omega2d}
\end{eqnarray}
in agreement with the argument presented in \cite{transverseDP}
where $\omega(d=2)$ should coincide with the Directed Polymer droplet exponent
$\omega_{DP}(D=d-1=1)=1/3$ \cite{Hus_Hen_Fis,Kar,Joh,Pra_Spo}.

\subsubsection{ Distribution of the logarithm of surface magnetization  }

\begin{figure}[htbp]
 \includegraphics[height=6cm]{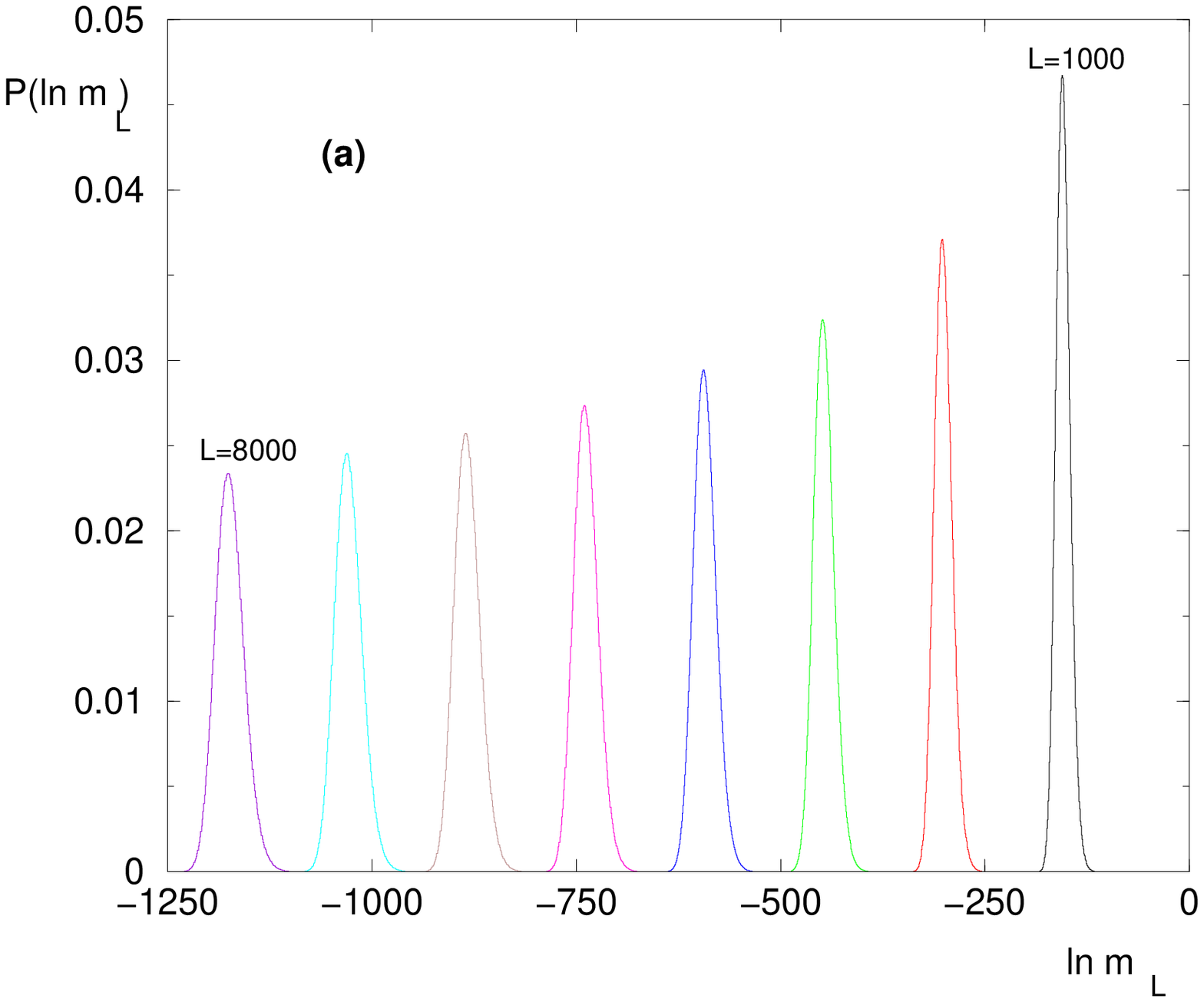}
\hspace{1cm}
 \includegraphics[height=6cm]{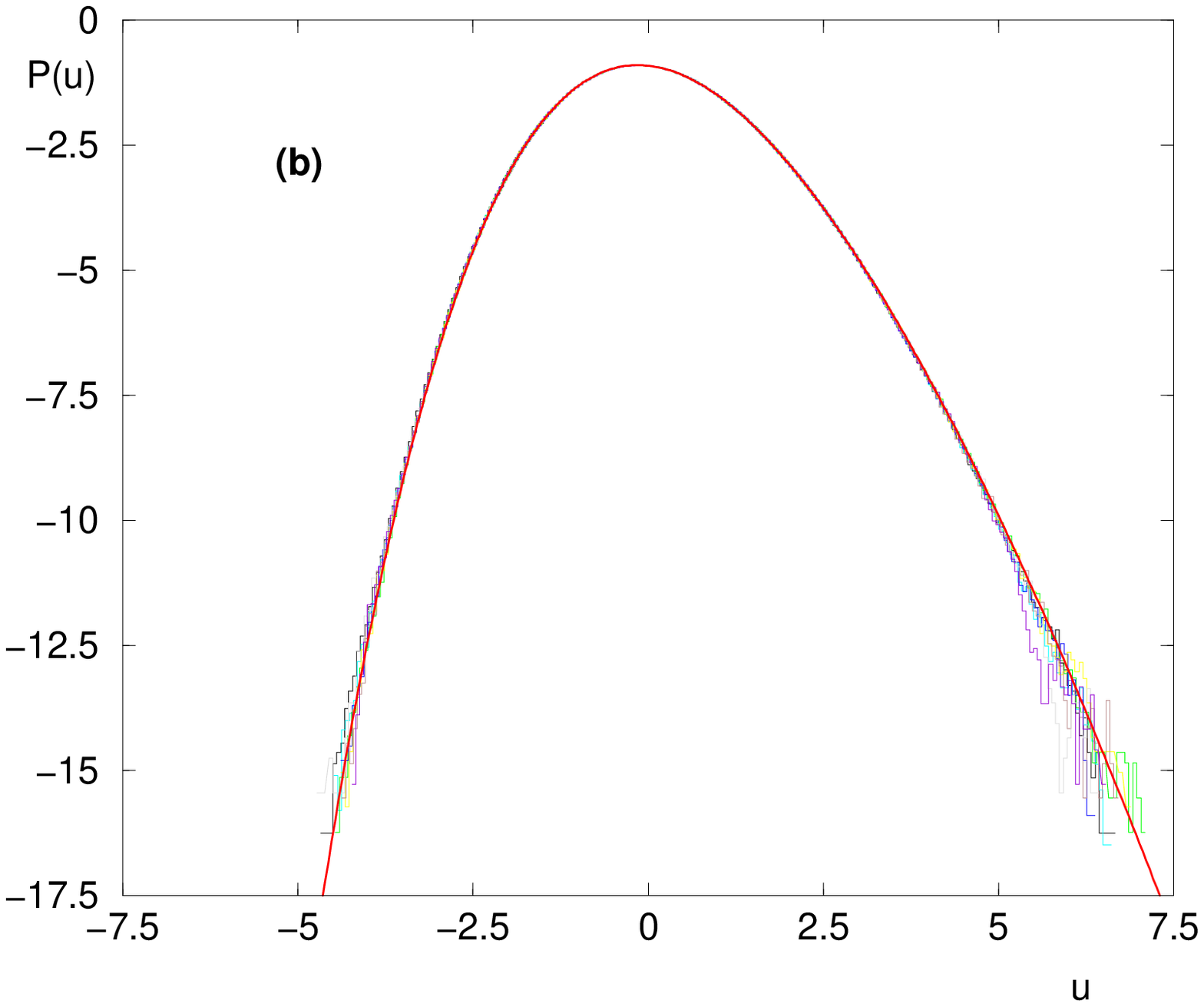}
\caption{ Disordered phase in $d=2$ (here $J=0.29$) :
(a) Evolution with $L$ of the probability distribution $P_L(\ln m_L^{surf})$
of the logarithm of surface magnetization : 
(b) Corresponding fixed distribution of the rescaled variable 
$u=(\ln m_L^{surf} -\ln m_L^{typ})/\Delta_L  $ in log-scale to show the tails,
compared to the exact Tracy-Widom GOE distribution (thick line).
  }
\label{fighistodisordered}
\end{figure}

We show on Fig \ref{fighistodisordered} (a) our numerical results
concerning histograms of the logarithm of the surface magnetization
in the disordered phase. Our conclusion is that the surface magnetization follows the scaling
\begin{eqnarray}
\ln (m_L^{surf}) \opsimeq_{L \to \infty} \ln (m_L^{typ})  +\Delta_L u
\label{msdisfull}
\end{eqnarray}
where the behaviors of the typical value 
$\ln (m_L^{typ})  \simeq -L/\xi_{typ}$ 
and of the width $\Delta_L \sim L^{\omega}$ have been already discussed 
above in Eqs \ref{msdis}  and \ref{deltaLomega}  respectively.
On Fig. \ref{fighistodisordered} (b), we show that the stable distribution $P(u)$
of the rescaled variable $u$ coincides with the GOE
 Tracy-Widom distribution, as expected from the correspondence with
the Directed Polymer model in the disordered phase \cite{transverseDP}.

\subsection{ Ordered phase }

\subsubsection{ Behavior of the typical surface magnetization $m_{\infty}^{typ}$
in the ordered phase $J>J_c$  }

\begin{figure}[htbp]
 \includegraphics[height=6cm]{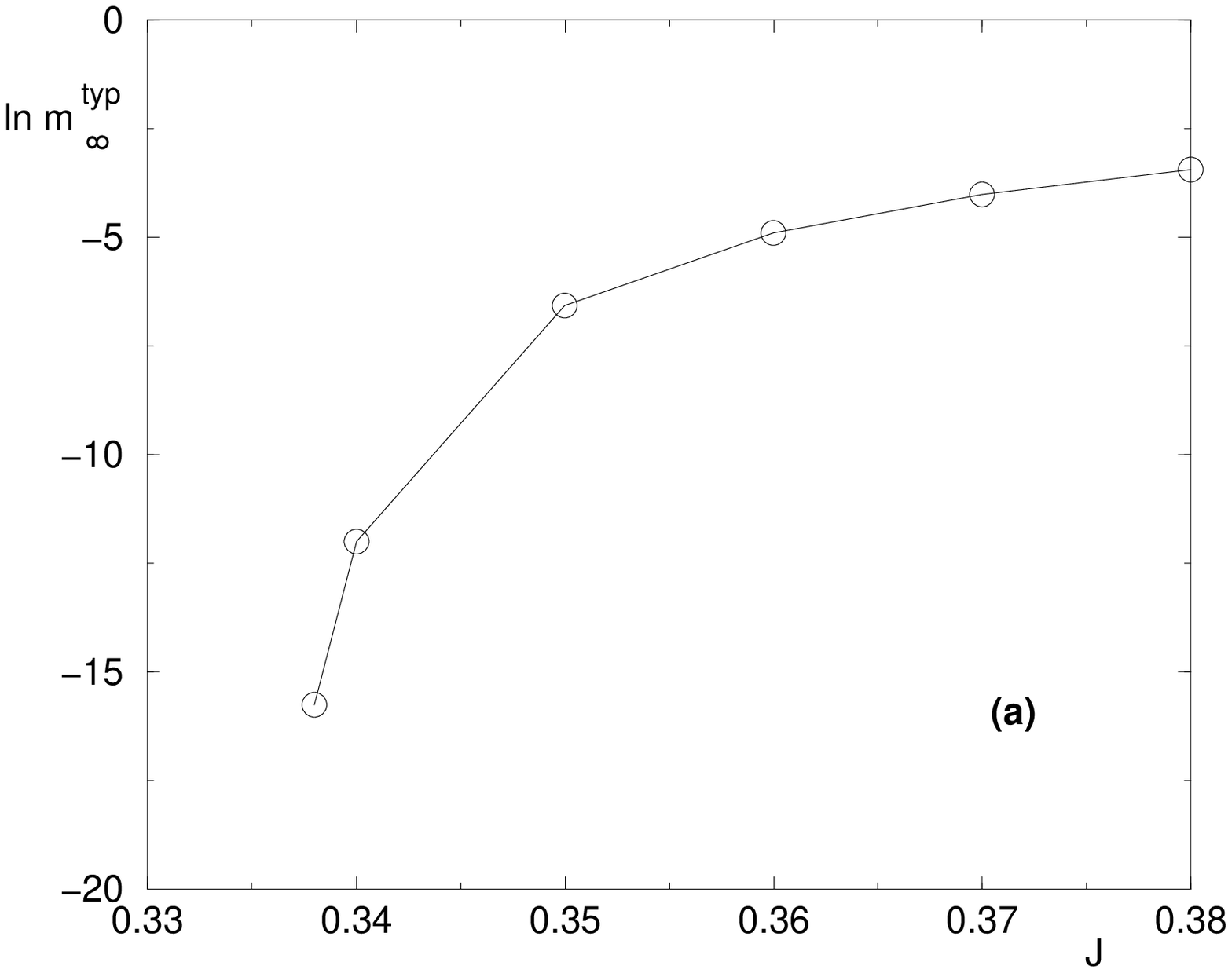}
\hspace{1cm}
\includegraphics[height=6cm]{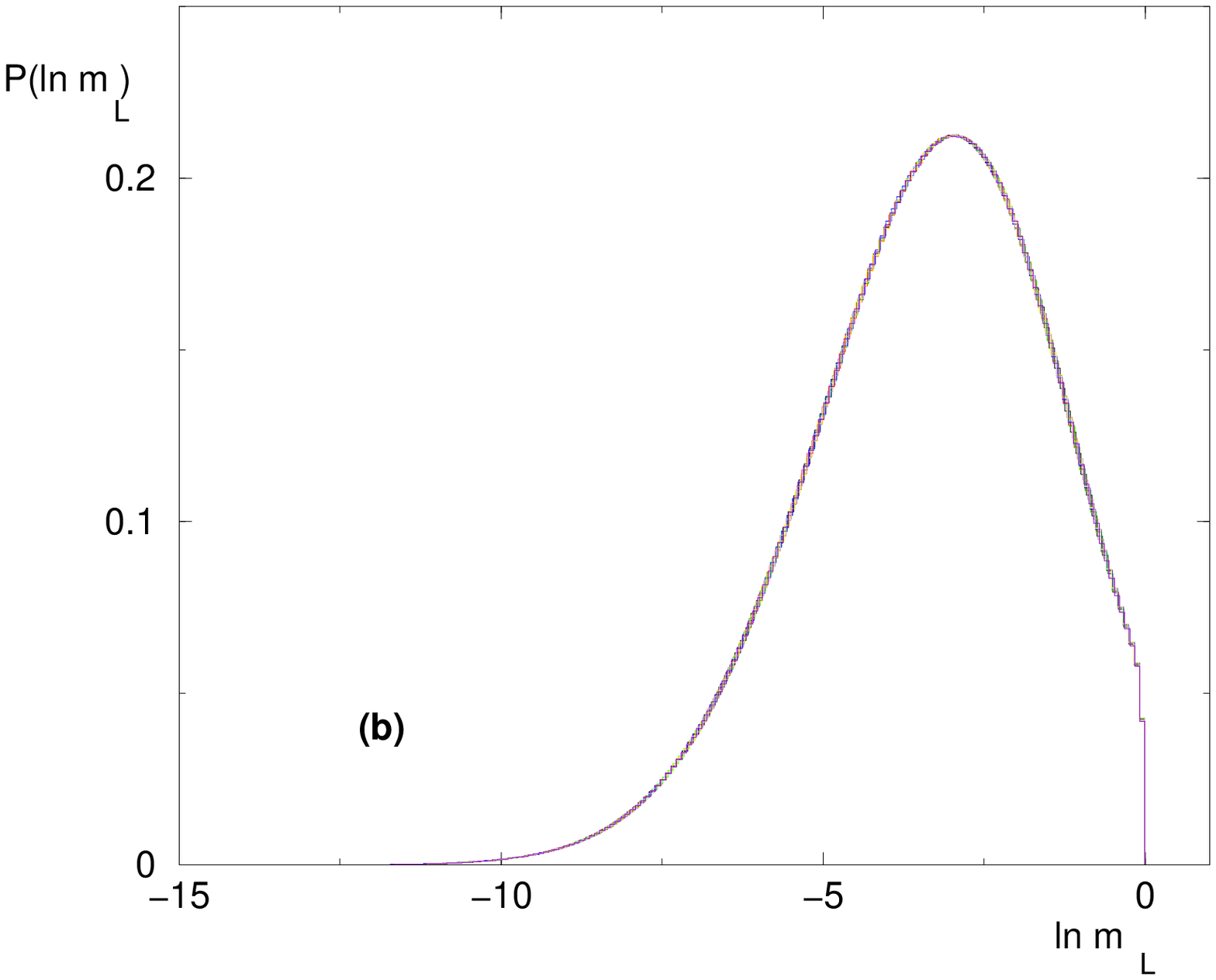}
\caption{ Ordered Phase $J>J_c$ in $d=2$ :
(a) 
Behavior of the asymptotic typical surface magnetization $m_{\infty}^{typ}$
as a function of the ferromagnetic coupling $J$ : our numerical data are
compatible with an essential singularity (Eq. \ref{defkappa})
of exponent $\kappa \simeq 0.5 $
(b) 
the probability distribution $P_L(\ln m_L^{surf})$ (here for $J=0.38$)
of the logarithm of surface magnetization remains fixed and attached at the origin
(as a consequence of the bound $m_L^{surf} \leq 1$).
  }
\label{fig2dordered}
\end{figure}

In the ordered phase, the typical surface magnetization remains finite 
in the limit where the number of generations $L$ diverges  
\begin{eqnarray}
 \ln m_L^{typ}(J>J_c)  \equiv \overline{ \ln m_L^{surf}(J>J_c) } \opsimeq_{L \to \infty} 
  \ln m_{\infty}(J>J_c)  > -\infty
\label{transdeloc}
\end{eqnarray}
and one expects an 
 essential singularity behavior 
\begin{eqnarray}
 \ln m_{\infty}^{typ}(J>J_c)  \oppropto_{J \to J_c^+} -  (J-J_c)^{- \kappa}   
\label{defkappa}
\end{eqnarray}
Our data shown on Fig. \ref{fig2dordered} (a)
can be fitted with the value
\begin{eqnarray}
  \kappa (d=2) \simeq 0.5
\label{kappa2d}
\end{eqnarray}
that can be related to other exponents via finite-size scaling
(see below around Eq. \ref{kappaomega})

\subsubsection{ Distribution of the logarithm of the surface magnetization  }

In the ordered phase, the probability distribution $P_L(\ln m_L^{surf})$
of the logarithm of surface magnetization remains fixed as $L$ varies,
and terminates discontinuously at the origin, 
as a consequence of the bound $m_L^{surf} \leq 1$ 
corresponding to $\ln m_L^{surf} \leq 0$ (see Fig. \ref{fig2dordered} (b))

\subsection{ Critical point }

\subsubsection{ Behavior of the typical surface magnetization $m_{L}^{typ}$ and of the width $\Delta_L$  }

\begin{figure}[htbp]
 \includegraphics[height=6cm]{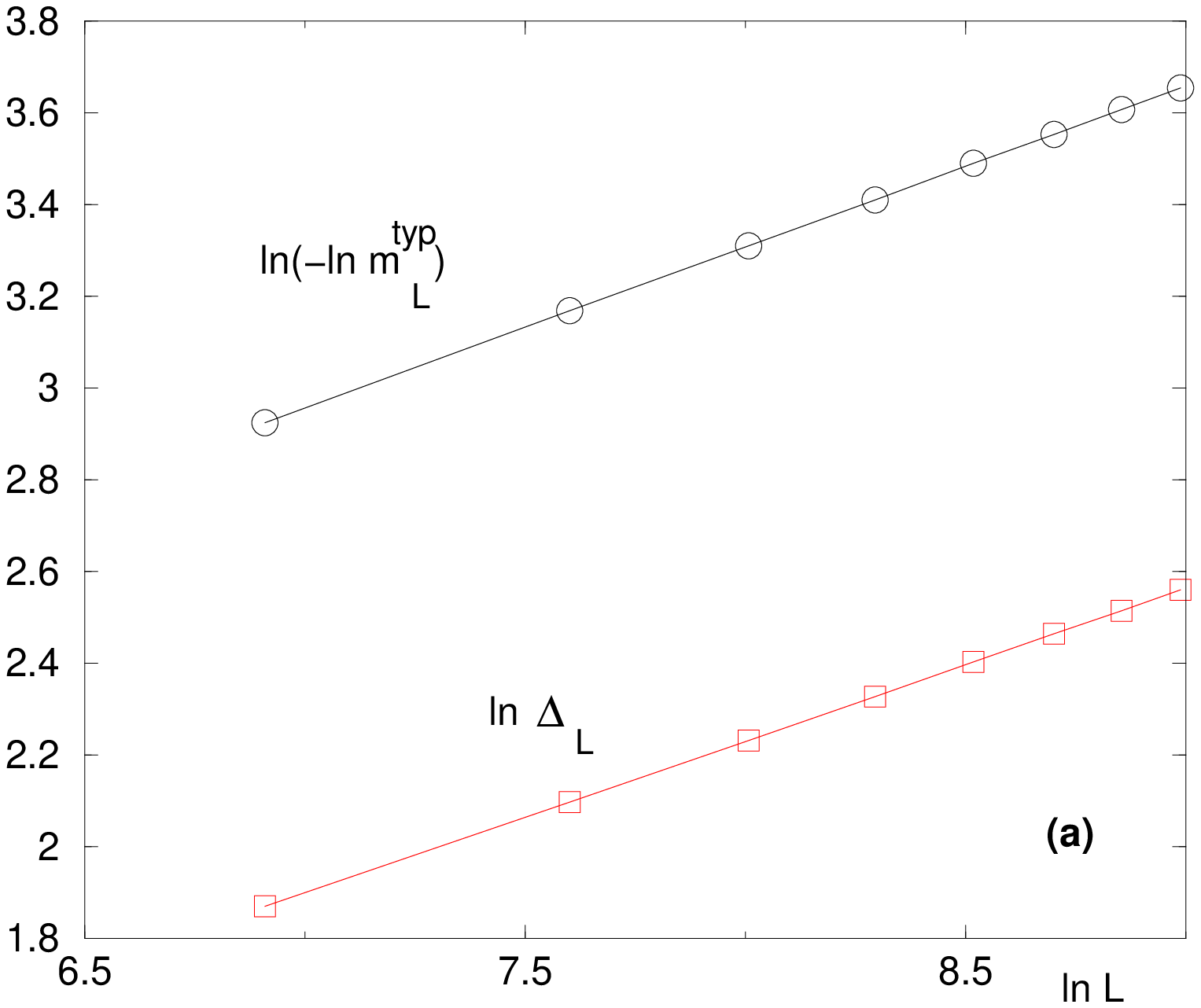}
\hspace{1cm}
 \includegraphics[height=6cm]{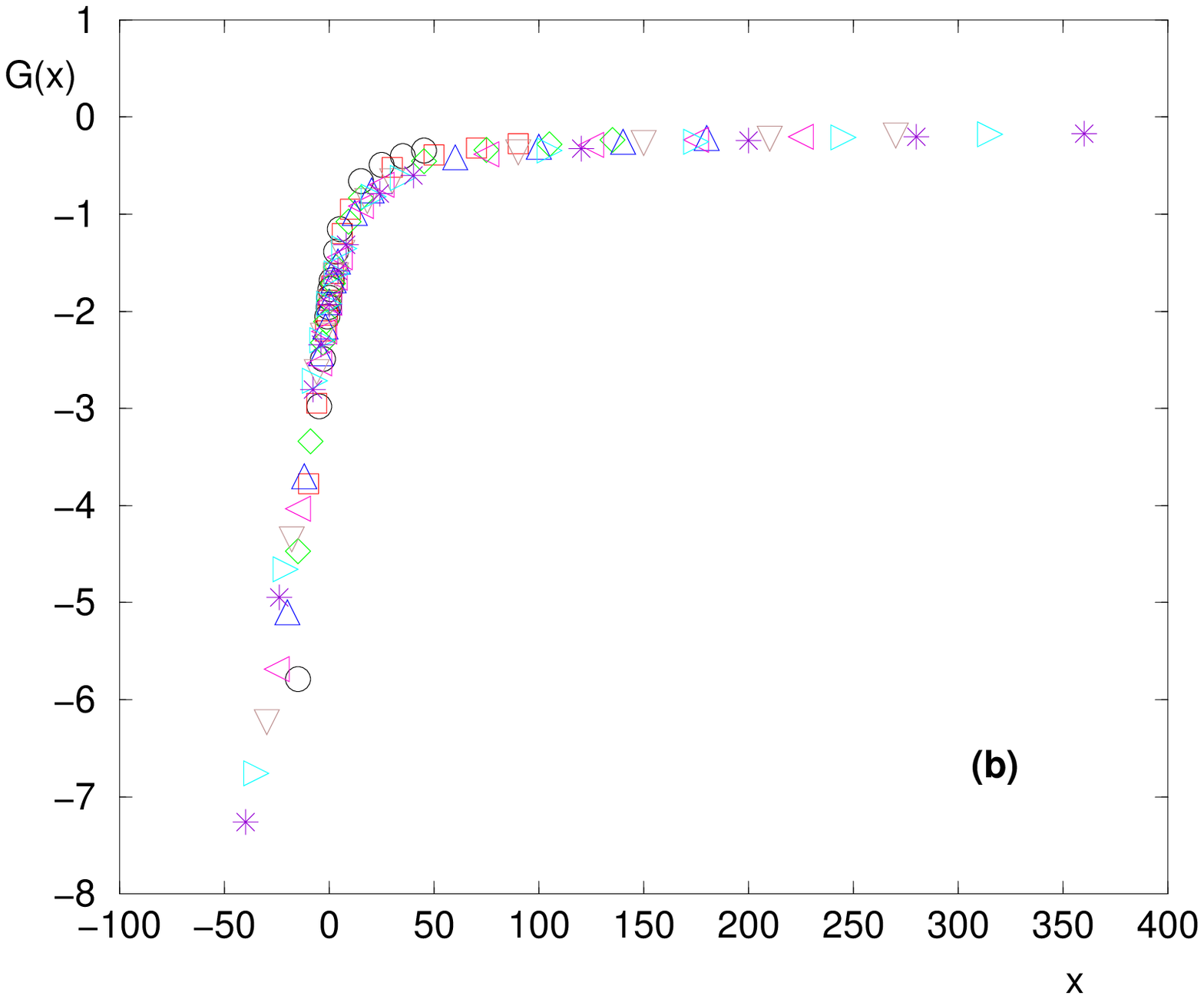}
\caption{ Critical point (here $J_c=0.335$) :
 (a)   Log-log plot 
of the logarithm of the typical surface magnetization $m_{L}^{typ}$ 
and of the width $\Delta_L$ : both slopes are of order $\omega_c \simeq 0.33$
 (see Eqs \ref{mscriti} and \ref{deltaLcriti})
 (b) Finite-size scaling of the typical surface magnetization
according to Eq. \ref{fssmtyp} 
with $J_c=0.335$, $\omega_c=0.33$ and $\nu_{av}=1.5$. }
\label{fig2dcriti}
\end{figure}

Exactly at criticality, one expects that the typical surface magnetization
follows an activated behavior of exponent $\omega_c<1$ (compare with Eq. \ref{msdis} in the disordered phase)
\begin{eqnarray}
\ln (m_L^{typ}(J=J_c)) \equiv  \overline{ \ln m_L^{surf}(J=J_c) }
\opsimeq_{L \to \infty} - L^{\omega_c}
\label{mscriti}
\end{eqnarray}
and that the width defined in Eq. \ref{defdeltaL} is also governed by the same exponent
\begin{eqnarray}
\Delta_L(J=J_c) \opsimeq_{L \to \infty} L^{\omega_c}
\label{deltaLcriti}
\end{eqnarray}
Our numerical data at $J_c \simeq 0.335$ shown on Fig. \ref{fig2dcriti} (a)
are compatible with these behaviors with the value
\begin{eqnarray}
\omega_c \simeq 0.33
\label{psicriti}
\end{eqnarray}
i.e. $\omega_c$ coincides with the fluctuation exponent $\omega$ 
measured in the disordered phase (see Eq. \ref{omega2d})

This last property implies that the finite-size scaling for the typical 
surface magnetization $m_{L}^{typ}$ involves some correlation length 
exponent $\nu_{av}$ different from $\nu_{typ}$
\begin{eqnarray}
\ln m_L^{typ}(J) \equiv \overline{ \ln m_L^{surf}(J)} \opsimeq - L^{\omega_c} G \left(x \equiv (J-J_c) L^{1/\nu_{av}}  \right)
\label{fssmtyp}
\end{eqnarray}
The matching with the behavior of Eq. \ref{msdisfull}
 in the disordered phase implies that 
\begin{eqnarray}
G(x) \oppropto_{x \to -\infty} (-x)^{\nu_{typ}}
\label{Gxneg}
\end{eqnarray}
and that $\nu_{av}$ reads
\begin{eqnarray}
\nu_{av} = \frac{\nu_{typ}}{1-\omega}
\label{nuavnutyp}
\end{eqnarray}
This relation can be understood within a rare events analysis 
for the averaged correlation in the disordered phase 
\cite{transverseDP}.
The values $\nu_{typ} \simeq 1$ and $\omega=1/3$ yield
\begin{eqnarray}
\nu_{av}(d=2) \simeq \frac{3}{2}
\label{nuav2d}
\end{eqnarray}
The matching of the finite-size scaling form of Eq. \ref{fssmtyp}
with the essential singularity of Eq. \ref{defkappa}
 in the ordered phase implies that
\begin{eqnarray}
G(x) \oppropto_{x \to + \infty} \frac{1}{x^{\kappa}}
\label{Gxpos}
\end{eqnarray}
with 
\begin{eqnarray}
\kappa=\omega_c \nu_{av} = \nu_{typ} \frac{\omega}{1-\omega}
\label{kappaomega}
\end{eqnarray}
The values $\nu_{typ}=1$ and $\omega=1/3$ yield
\begin{eqnarray}
\kappa(d=2) \simeq \frac{1}{2}
\label{kappa2dtheo}
\end{eqnarray}
in agreement with the estimate of Eq. \ref{kappa2d}.

As shown on Fig \ref{fig2dcriti} (b), our numerical data collapse well
with the finite-size scaling form of Eq. \ref{fssmtyp} with $\nu_{av}=1.5$.

\subsubsection{ Distribution of the logarithm of surface magnetization  }

\begin{figure}[htbp]
 \includegraphics[height=6cm]{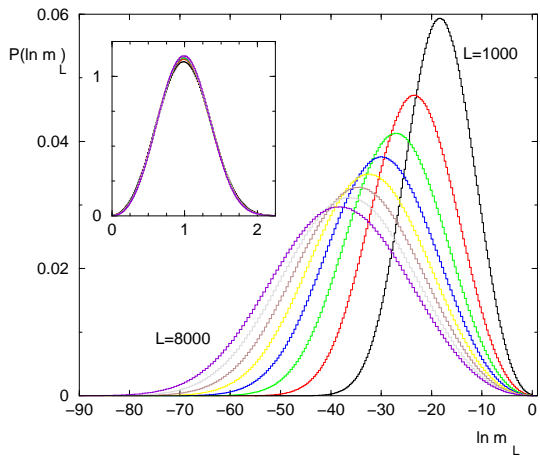}
 \hspace{1cm}
\caption{ Critical point in $d=2$ (here $J_c=0.335$) :
 Evolution with $L$ of the probability distribution $P_L(\ln m_L^{surf})$
of the logarithm of surface magnetization. 
Inset :  Corresponding fixed distribution of the rescaled variable 
$v=(\ln m_L^{surf})/\ln m_L^{typ} $. 
  }
\label{fighistocriti}
\end{figure}

At criticality, the rescaled variable
\begin{eqnarray}
v \equiv \frac{\ln m_L^{surf}}{\ln m_L^{typ}} 
\propto - \frac{\ln m_L^{surf}}{L^{\omega_c}} 
\label{mscritihisto}
\end{eqnarray}
 remains a positive random variable of order $O(1)$ as $L \to +\infty$.
Our numerical measure of its probability distribution $P(v)$ shown on Fig. \ref{fighistocriti} 
is compatible with a power-law singularity near the origin
\begin{eqnarray}
P(v) \opsimeq_{v \to 0^+} v^a
\label{pvcriti}
\end{eqnarray}
with an exponent of order $a \geq 2$ that we do not measure precisely.
Note that this is different from the case $d=1$ where $P(v=0)$ is finite ($a=0$).
We have not been able to find a physical argument to predict the value of $a$ in $d=2$.
This exponent $a$ will directly influence the scaling of
the moments of the surface magnetization, as we now discuss.

\subsubsection{ Moments of the surface magnetization  }

In contrast to the activated behavior
 of the typical surface magnetization $m_{L}^{typ}$
of Eq. \ref{mscriti}, the moments of the surface magnetization 
are expected to follow a power-law, as a consequence of 
the following rare events analysis :
the surface magnetization of Eq. \ref{mscritihisto}
will be of order $O(1)$ if the random variable $v$ happens
to be smaller than $1/L^{\omega_c}$. Taking into account the behavior
of Eq. \ref{pvcriti}, this will happen with probability
\begin{eqnarray}
Prob(m_L^{surf}=1) \simeq \int_0^{1/L^{\omega_c}} dv P(v) \sim \int_0^{1/L^{\omega_c}} dv v^a
\oppropto_{L \to \infty} L^{- \omega_c (1+a) } 
\label{prob1xs}
\end{eqnarray}
and all moments will be governed by this power-law 
\begin{eqnarray}
\overline{(m_L^{surf})^k} \simeq Prob(m_L^{surf}=1)
\oppropto_{L \to \infty} L^{- x_s } \ \ {\rm with } \ \ x_s=\omega_c (1+a)
\label{msavcriti}
\end{eqnarray}
independently of the order $k$.

Our numerical data for the three first moments $k=1,2,3$
and various sizes are compatible with Eq. \ref{msavcriti}
with an exponent
\begin{eqnarray}
x_s (d=2) \simeq 1.2
\label{xs2d}
\end{eqnarray}
The relation of Eq. \ref{msavcriti} then corresponds to
\begin{eqnarray}
a(d=2) \simeq 2.6
\label{a2d}
\end{eqnarray}

In the ordered phase, our numerical data are compatible with the power-law
\begin{eqnarray}
\overline{ m_L^{surf} } \propto (J-J_c)^{\beta_s}
\label{defbetas}
\end{eqnarray}
with 
\begin{eqnarray}
\beta_s (d=2) = x_s \nu_{av} \simeq 1.8 
\label{beta2d}
\end{eqnarray}

\section{ Numerical results in dimension $d=3$  }

\label{sec_dim3}

In this section, we present the numerical results obtained 
with the following sizes $L$ and the corresponding numbers $n_s(L)$ of disordered samples
of volume $L^3$
\begin{eqnarray}
L && = 10^2, 2. 10^2 , 3.10^2 , 4.10^2 , 5.10^2 , 6.10^2, 7.10^2, 8. 10^2 \nonumber \\
n_s(L) && = 27.10^3, 7.10^3, 3.10^3, 16.10^2, 10^3, 7.10^2, 5.10^2, 4.10^2
\label{numed3}
\end{eqnarray}
For each sample $\alpha$, we collect the $L^2$ values of the surface
 magnetization 
at the different points  of the surface.
Average values and histograms are then based on these $L^2 \times n_s(L)$ values.
We consider again the disorder distribution of Eq. \ref{lognormal}
and take $J$ as the control parameter of the transition.

\subsection{ Disordered phase }

\begin{figure}[htbp]
 \includegraphics[height=6cm]{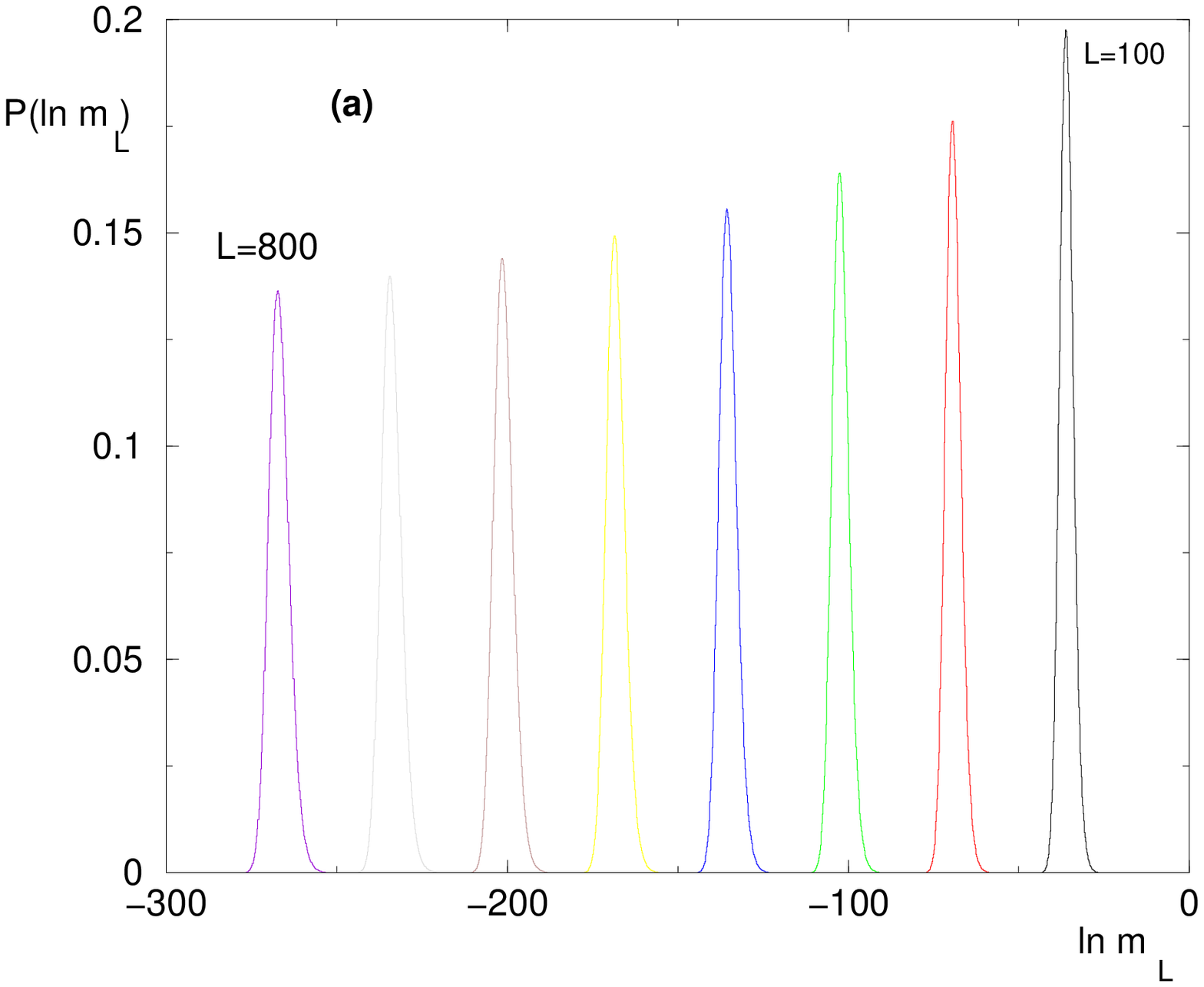}
\hspace{1cm}
 \includegraphics[height=6cm]{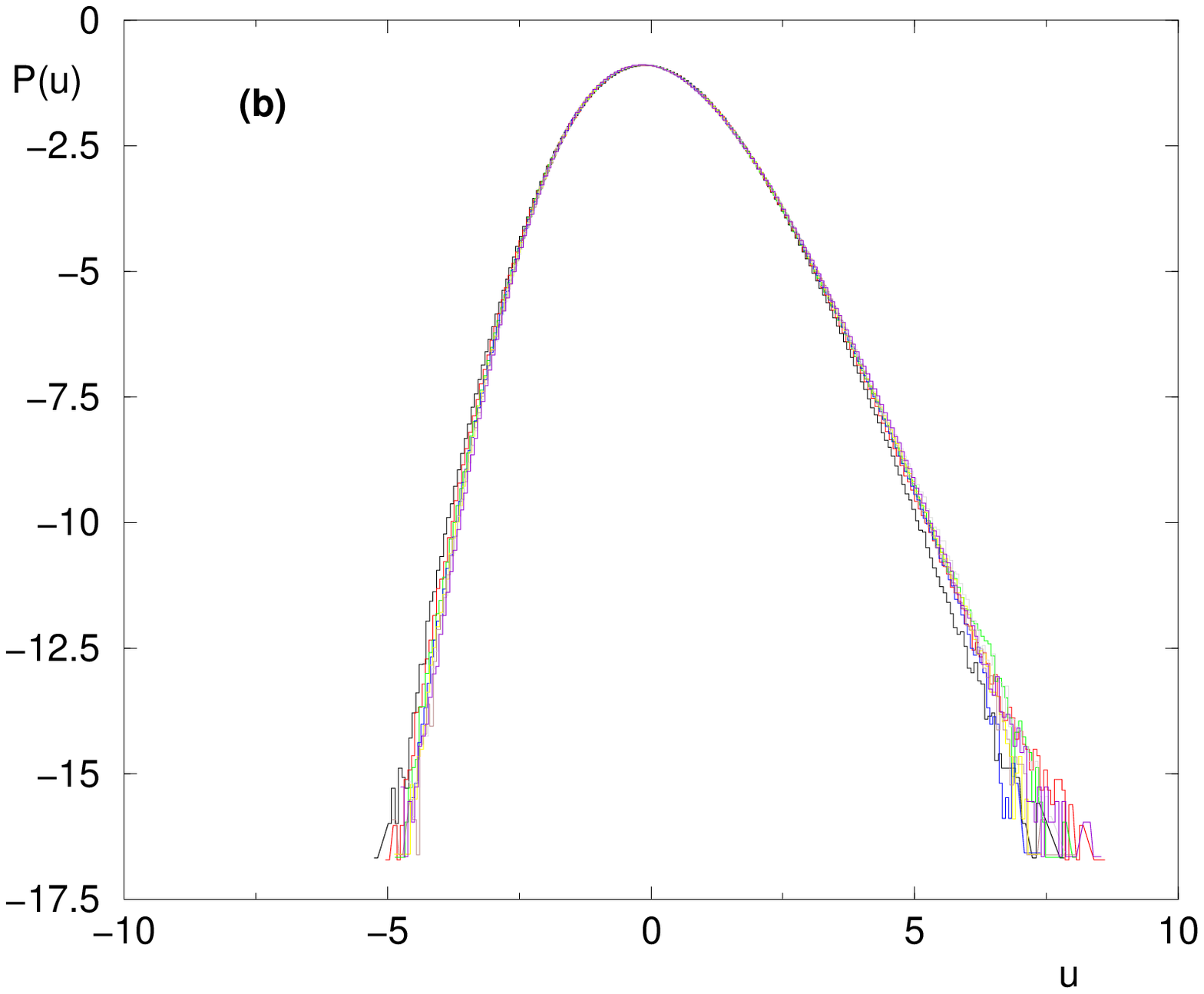}
\caption{ Disordered phase in $d=3$ (here $J=0.11$) :
(a) Evolution with $L$ of the probability distribution $P_L(\ln m_L^{surf})$
of the logarithm of surface magnetization : 
(b) Corresponding fixed distribution of the rescaled variable 
$u=(\ln m_L^{surf} -\ln m_L^{typ})/\Delta_L  $ in log-scale to show the
 tails.
  }
\label{fighistodisordered3d}
\end{figure}

Our data follow the scaling of Eq. \ref{msdisfull},
with the following properties :

(i) the scaling of the typical surface magnetization is given by
 Eq \ref{msdis}, and the typical correlation length exponent of Eq. \ref{xityp},
 seems again very close to unity
\begin{eqnarray}
\nu_{typ} \simeq 1
\label{nutyp3d}
\end{eqnarray}

(ii) the width $\Delta_L$ of Eq. \ref{defdeltaL} grows as the power-law
 of Eq. \ref{deltaLomega}
with the exponent
\begin{eqnarray}
\omega(d=3) \simeq 0.24
\label{omega3d}
\end{eqnarray}
that coincides with the numerical values of
 the droplet exponent of the Directed Polymer model with $D=d-1=2$
transverse dimensions \cite{Tan_For_Wol,Ala_etal,perlsman,KimetAla,Mar_etal,DPtails,schwartz},
in agreement with the argument presented in \cite{transverseDP}.

(iii) As $L$ grows, the evolution of the probability distribution $P_L(\ln m_L^{surf})$
is shown on Fig. \ref{fighistodisordered3d} (a).
The corresponding fixed distribution of the rescaled random variable 
$u= (\ln m_L^{surf} -\ln m_L^{typ})/\Delta_L $ is shown 
 on Fig \ref{fighistodisordered3d} (b).

\subsection{ Critical point  }

\begin{figure}[htbp]
 \includegraphics[height=6cm]{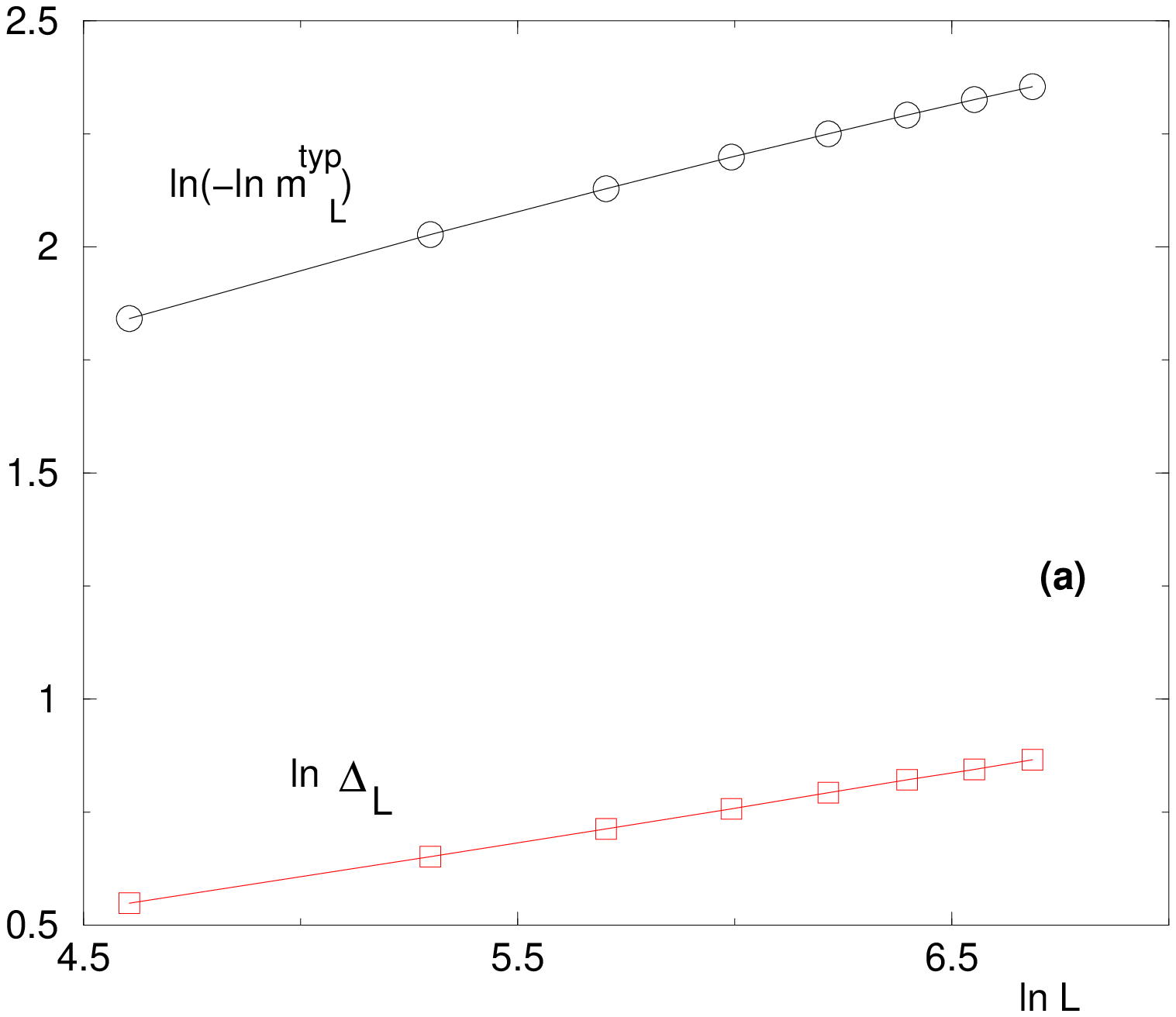}
\hspace{1cm}
 \includegraphics[height=6cm]{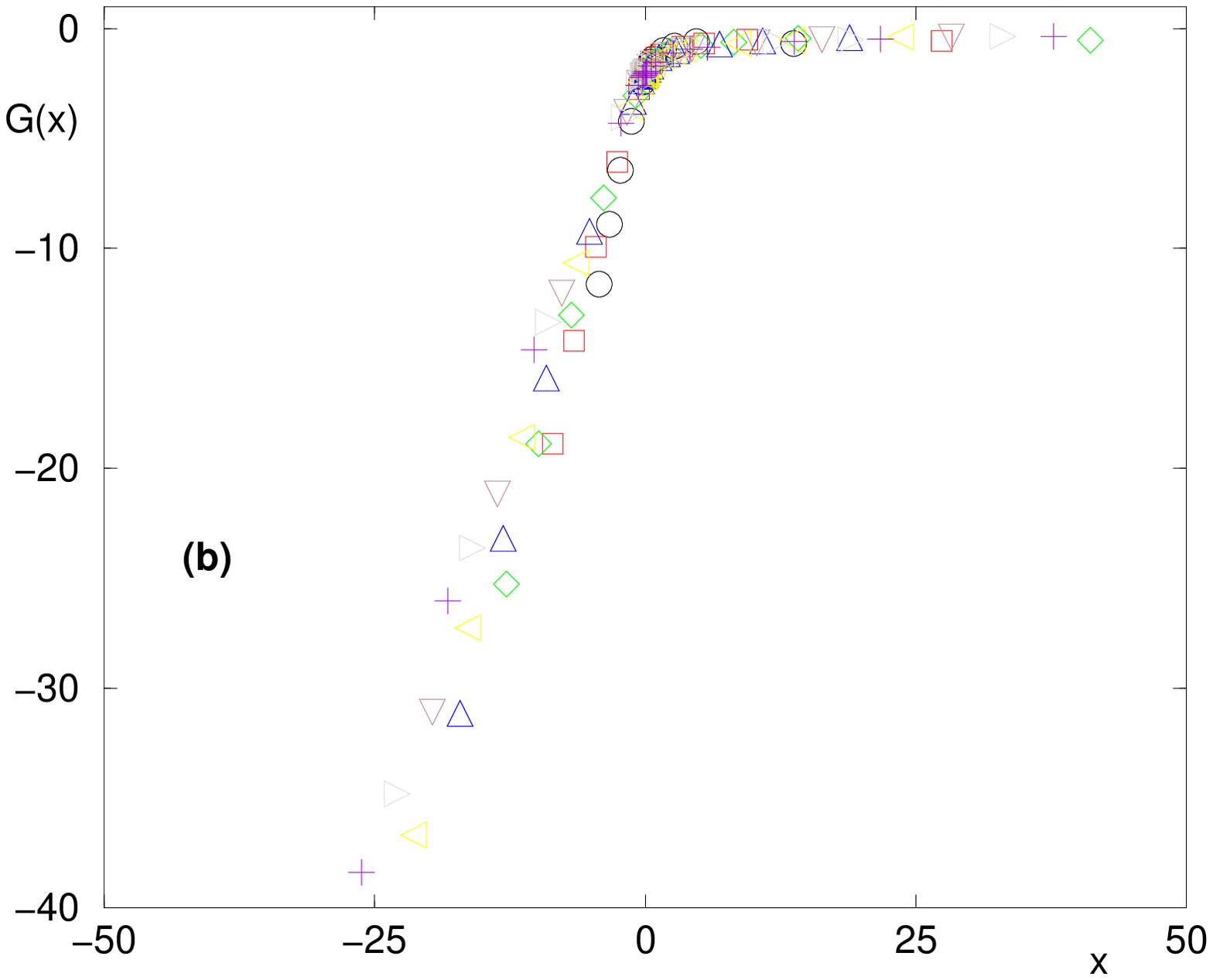}
\caption{ Typical surface magnetization in the critical region in $d=3$
(a) Log-log plot of the logarithm of the typical surface magnetization $m_{L}^{typ}$ 
and of the width $\Delta_L$  at criticality  $J_c=0.1528$ : 
both slopes are of order $\omega_c \simeq 0.24$ (see Eqs \ref{mscriti} and \ref{deltaLcriti} )
(b) Finite-size scaling of the typical surface magnetization in $d=3$
according to Eq. \ref{fssmtyp} with  $\nu_{av}=1.32$.
  }
\label{figfss3d}
\end{figure}

At criticality $J_c=0.1528$, we find that the exponent $\omega_c$ of Eqs \ref{mscriti}
and \ref{deltaLcriti} coincides with the value of $\omega$ of Eq. \ref{omega3d}
concerning the disordered phase (see Fig. \ref{figfss3d} (a))
\begin{eqnarray}
\omega_c(d=3) \simeq 0.24
\label{omegas3d}
\end{eqnarray}

We show on Fig. \ref{figfss3d} (b) the finite-size scaling analysis
of Eq. \ref {fssmtyp} for the logarithm of the typical surface magnetization
with the averaged correlation length exponent $\nu_{av}=1/(1-\omega) \simeq 1.32$.

\begin{figure}[htbp]
 \includegraphics[height=6cm]{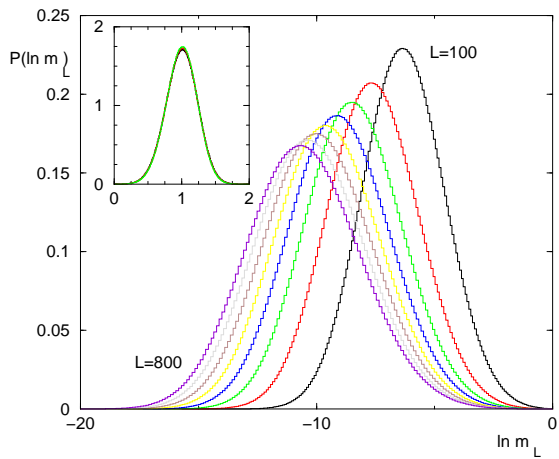}
\hspace{1cm}
\caption{ Critical point in $d=3$ (here $J_c=0.1528$) :
 Evolution with $L$ of the probability distribution $P_L(\ln m_L^{surf})$
of the logarithm of surface magnetization.
 Inset :  Corresponding fixed distribution of the rescaled variable 
$v=(\ln m_L^{surf})/\ln m_L^{typ} $. 
  }
\label{fighistocriti3d}
\end{figure}

We show on Fig. \ref{fighistocriti3d}
our numerical data for the probability distribution of the surface magnetization
at criticality~: the fixed point distribution $P(v)$
of the rescaled variable $v$ of Eq. \ref{mscritihisto}
displays a power-law singularity near the origin
 (Eq \ref{pvcriti}), that will determine the scaling of all moments 
of the surface magnetization according to Eq. \ref{msavcriti}.
Our numerical data for the moments are compatible with Eq. \ref{msavcriti}
with an exponent
\begin{eqnarray}
x_s (d=3) \simeq 1.34
\label{xs3d}
\end{eqnarray}
that would correspond to
\begin{eqnarray}
a(d=3) = \frac{x_s}{\omega_c} -1 \simeq 4.5
\label{a3d}
\end{eqnarray}
and to the exponent (Eq. \ref{defbetas})
\begin{eqnarray}
\beta_s (d=3) = x_s \nu_{av} \simeq 1.76
\label{beta3d}
\end{eqnarray}

\section{ Conclusion  }

\label{sec_conclusion}

Since the 'Cavity-Mean-Field' approximation developed for the Random Transverse Field Ising Model
on the Cayley tree \cite{ioffe,feigelman,dimitrova} has been found
 to reproduce the known exact result for the surface magnetization in $d=1$ \cite{dimitrova},
we have proposed to extend these ideas in finite dimensions $d>1$
via a non-linear transfer approach for the surface magnetization.
In the disordered phase, the linearization (Eq \ref{milinear}) of the transfer equations 
correspond to the transfer matrix for a Directed Polymer in a random medium
of transverse dimension $D=d-1$, in agreement with the leading order perturbative
scaling analysis \cite{transverseDP}. 

We have presented numerical results of this non-linear transfer approach in dimensions $d=2$
and $d=3$, where large system sizes can be easily studied.
 In both cases, we have found that the critical point is
governed by Infinite Disorder scaling. In particular exactly at criticality,
the one-point surface magnetization scales as $\ln m_L^{surf} \simeq - L^{\omega_c} v$,
where $\omega_c(d)$ coincides with the droplet exponent $\omega_{DP}(D=d-1)$
of the corresponding Directed Polymer model, with
 $\omega_c(d=2)=1/3$ and $\omega_c(d=3) \simeq 0.24$. The distribution $P(v)$ 
of the positive random variable $v$ of order $O(1)$
presents a power-law singularity near the origin $P(v) \propto v^a$ with $a(d=2,3)>0$ so that all
moments of the surface magnetization are governed by the same power-law decay
$\overline{ (m_L^{surf})^k } \propto L^{- x_s}$ with $x_s=\omega_c (1+a)$ 
independently of the order $k$.
Our conclusion is thus that this non-linear transfer approach is able
to lead to Infinite Disorder scaling, that had been found previously
via Monte-Carlo in $d=2$ \cite{pich,rieger} and via Strong Disorder RG 
in $d=2,3,4$ \cite{motrunich,fisherreview,lin,karevski,lin07,yu,kovacsstrip,
kovacs2d,kovacs3d,kovacsentropy,kovacsreview}.
Exactly at criticality, the presence of activated scaling $\ln m_L^{surf} \simeq - L^{\omega_c} v$
means that the linearization of Eq. \ref{milinear} is typically still valid 
also at criticality (and not only in the disordered phase),
so that the identity $\omega_c=\omega_{DP}(D=d-1)$ can be understood.
The rare cases where this linearization is not valid at criticality is when
the positive random variable $v$ happens to be smaller than $1/L^{\omega_c}$.
Our conclusion is thus the following :

(i) in the disordered phase and for 'typical' situations exactly at criticality,
 the linearization of Eq. \ref{milinear} is valid
and coincides with the leading order perturbative
scaling analysis \cite{transverseDP} : it is thus
 likely to give exact values for critical exponents,
in particular for the exponent $\omega_c$ of activated scaling.

(ii) in the ordered phase and for 'rare' situations at criticality,
 the non-linear terms of the transfer approach plays an important role.
Since they come from an uncontrolled approximation, the critical exponents
like $\beta_s$ and $x_s$ that are determined by these non-linear contributions
could be different from the exact ones. 
To judge the accuracy of this approximation, it would be very helpful to compare
with other approaches like Quantum Monte-Carlo and Strong Disorder RG
(but up to now, these other approaches have not studied surface properties).

\end{document}